\documentclass[amssymb,prb,twocolumn,showpacs]{revtex4}
\usepackage{amsmath,amsthm,amsfonts}
\usepackage{epsf}
\usepackage{amssymb}
\usepackage{graphicx}

\begin{document}
\title{Charge qubit dynamics in a double quantum dot coupled to phonons}
\author{Zhuo-Jie Wu}\email{wuzj@sjtu.edu.cn}
\author{Ka-Di Zhu} \author{Xiao-Zhong Yuan} \author{Yi-Wen Jiang} \author{Hang
Zheng}\affiliation{Department of Physics, Shanghai Jiao Tong
University, Shanghai 200240, China}
\date{\today{ }}

\begin{abstract}
The dynamics of charge qubit in a double quantum dot coupled to
phonons is investigated theoretically in terms of a perturbation
treatment based on a unitary transformation. The dynamical
tunneling current is obtained explicitly. The result is compared
with the standard perturbation theory at Born-Markov
approximation. The decoherence induced by acoustic phonons is
analyzed at length. It is shown that the contribution from
deformation potential coupling is comparable to that from
piezoelectric coupling in small dot size and large tunneling rate
case. A possible decoupling mechanism is predicted.
\end{abstract}
 \maketitle
\section{Introduction}
Computers based on quantum mechanics is proven, by computational
theory, more efficient at some specific calculations than those
based on classical physics.\cite{1,2,3} The first step to build a
quantum computer is the realization of its building block, quantum
bit (qubit). Within the last decade, various schemes have been
proposed and many of them have even been realized.\cite{4,5,6,7,8}
Among them, the electrically controlled charge qubit in a
semiconductor double quantum dot has the potential advantages of
being arbitrarily scalable to large system and compatible with the
current microelectronics technology. Besides, the double-dot
system is also extremely useful in basic physics as it enables us
to investigate the decoherence and dissipation of a small quantum
system interacting with its environment.

Various designs of double-dot qubits have been studied in
experiment.\cite{9,10,11,12} Recently, Hayashi et al. successfully
realized coherent manipulation of electronic states in a
double-dot system implemented in a GaAs/AlGaAs heterostructure
containing a two dimensional electron gas. The damped Rabi
oscillation was observed in the time domain in their experiment
and an experienced formula was presented to fit the experimental
data.\cite{13} Fujisawa et al. explained the transport in this
system through density matrix simulations.\cite{14} But the phonon
effect was not included. However, the phonon induced decoherence
is significant according to the analysis of Ref. [13]. Considering
the interaction with the piezoelectric acoustic phonons, Brandes
et al. investigated in detail the single electron tunnelling in a
double quantum dot.\cite{15,16,17,18,19} Fedichkin and Fedorov
also studied the error rate of the charge qubit coupled to an
acoustic phonon reservoir. \cite{20,21} But no dynamical tunneling
current is presented by all these work. In a latest review
paper,\cite{22} Brandes derived the dynamical tunneling current in
the weak electron-phonon dissipation limit through Born-Markov
approximation (PER). But the result is rather complicated.

In this work, we study the damped Rabi oscillation observed in
Ref. [13]. The quantum dynamics of a single electron tunneling in
the double-dot system is investigate without applying the
Born-Markov approximation to the electron-phonon interaction. A
simple explicit expression of dynamical tunneling current is
presented through perturbation treatment based on a unitary
transformation. The phonon (both the deformation potential and
piezoelectric contribution are included) induced decoherence is
investigated at length and possible decoupling mechanism is
presented.

This paper is organized as follows. In Sec. II we introduce the
model Hamiltonian for the charge qubit and solve it in what
follows. A comparison with PER approach is given in Sec. III. In
Sec. IV, we derive the spectral density of a double quantum dot
and then analyze the phonon induced decoherence. The conclusions
is given in the Sec. V at last.

\section{Model and theory}
In this section, we shall introduce the model Hamiltonain for the
double quantum dot and then develop general theory leading to
explicit expression for the dynamical current. We will focus on
the phonon effect on qubit dynamics.

\subsection{Model Hamiltonian}
The double quantum dot consists of left and right dots connected
through an interdot tunneling barrier. Due to Coulomb blockade,
only one excess electron is allowed to occupy the left or right
dot, which defines two basis vectors of $|L\rangle$ and
$|R\rangle$ (with the energy level $\varepsilon_{L}$ and
$\varepsilon_{R}$ respectively) in the Hilbert space. The energy
difference between these two states
$\varepsilon=\varepsilon_{L}-\varepsilon_{R}$ can be controlled by
the source-drain voltage $V_{sd}$.\cite{13} Considering the
coupling to its environment, the double dot can be described by
the Hamiltonian:
\begin{eqnarray}
H=H_e+H_p+H_{ep}+H_r+H_{er}.
\end{eqnarray}
Here the qubit Hamiltonian reads($\hbar=1$)
\begin{eqnarray}
H_e=-\frac{1}{2}\varepsilon(t)\sigma_z+T_c\sigma_x,
\end{eqnarray}
where $T_c$ is the interdot tunneling, $\sigma_x$ and $\sigma_z$
are Pauli matrix with
$\sigma_z=|L\rangle{\langle}L|-|R\rangle{\langle}R|$ and
$\sigma_x=|L\rangle{\langle}R|+|R\rangle{\langle}L|$. If the qubit
isolates from any other degrees of freedom, the excess electron
would oscillate coherently between two dots with the Rabi
frequency $\Delta=\sqrt{\varepsilon^2+4T^2_c}$. However, the qubit
has to couple to its environment (phonons and electron reservoir
in leads) in practice. $H_p$ and $H_{ep}$ stand for the phonon
reservoir and its coupling to charge qubit respectively. They can
be written as follows:
\begin{eqnarray}
H_p&=&\sum_{\bf{q},\lambda}\omega_{\bf{q},\lambda}b^{\dag}_{\bf{q},\lambda}b_{\bf{q},\lambda},\\
H_{ep}&=&\frac{1}{2}\sigma_z\sum_{\bf{q},\lambda}(M_{\bf{q},\lambda}b^{\dag}_{\bf{q},\lambda}
+M^*_{\bf{q},\lambda}b_{\bf{q},\lambda}),
\end{eqnarray}
where $b^{\dag}_{\bf{q},\lambda}$ ($b_{\bf{q},\lambda}$) and
$\omega_{\bf{q},\lambda}$ are the creation (annihilation)
operators and energy of the phonons with the wave vector $\bf{q}$
and polarization $\lambda$, $M_{\bf{q},\lambda}$ is the
electron-phonon coupling constant. The effects of the phonon bath
are fully described by a spectral density
\begin{eqnarray}
J(\omega)=\sum_{\bf{q},\lambda}|M_{\bf{q},\lambda}|^2\delta(\omega-\omega_{\bf{q},\lambda}).
\end{eqnarray}
$H_r$ and $H_{er}$ in the Hamiltonian $H$ stand for the electron
reservoir in leads and its coupling to charge qubit respectively.

In experiment, a pulse technique is used to switch the the
$V_{sd}$ from large bias in the initialization process (an excess
electron localizes in the left dot) to the zero bias in the
manipulation process (the double dot is isolated from leads, and
the excess electron tunnels resonantly (i.e., $\varepsilon=0$)
back and forth between two dots).\cite{13} Restoring a large bias
voltage $V_{sd}$ after the pulse time $t$ gives the measurement of
dynamical elastic tunneling current which stands for the
probability $n(t)$ of the excess electron in the right dot at that
exact time.

Neglecting the higher order tunneling (cotunneling) between leads
and the dots, the effective Hamiltonian in the manipulation
process reads:
\begin{eqnarray}
H_{eff}=T_c\sigma_x+\sum_{\bf{q}}\omega_{\bf{q}}b^{\dag}_{\bf{q}}}b_{{\bf{q}}}+
\frac{1}{2}\sigma_z\sum_{\bf{q}}(M_{\bf{q}}b^{\dag}_{\bf{q}}+M^{\ast}_{\bf{q}}{b_{\bf{q}}).
\end{eqnarray}
Here, for the sake of simplicity, we omit the polarization, since
it makes no difference in the theory below. When it makes
difference (in Sec. IV), it will be included again and noted out
explicitly. This effective Hamiltonian is the starting point for
our theory.

\subsection{Theory}
The effective Hamiltonian $H_{eff}$ is equivalent to the
spin-boson Hamiltonian in zero bias case. Though it seems rather
simple, it cannot be solved exactly. Various analytical or
numerical approaches have been proposed to obtain an approximate
solution to it.\cite{23,24}

Here, we apply a canonical transformation: $ H^{'}=
\exp(S)H_{eff}\exp(-S)$ , with the generator:\cite{25,26,wuzj}
\begin{equation}
S=\sum_{\bf{q}}\frac{\xi_{\bf{q}}}{2\omega_{\bf{q}}}(M_{\bf{q}}b^{\dag}_{\bf{q}}-M^{\ast}_{\bf{q}}b_{\bf{q}})\sigma_{z}
.
\end{equation}
Thus we get the Hamiltonian $H^{'}$, and decompose it into
$H^{'}=H^{'}_{0}+H^{'}_{1}+H^{'}_{2}$, where
\begin{eqnarray}
H^{'}_{0}&=&{\eta}T_c\sigma_{x}+\sum_{\bf{q}}\omega_{\bf{q}}b^{\dag}_{\bf{q}}b_{{\bf{q}}}
-\sum_{\bf{q}}\frac{|M_{\bf{q}}|^{2}}{4\omega_{\bf{q}}}\xi_{\bf{q}}(2-\xi_{\bf{q}}),\\
H^{'}_{1}&=&\frac{1}{2}\sigma_{z}\sum_{\bf{q}}(1-\xi_{\bf{q}})(M_{\bf{q}}b^{\dag}_{\bf{q}}+M^{\ast}_{\bf{q}}{b_{\bf{q}}})+
\eta{T_c}i\sigma_{y}A,
\\
H^{'}_{2}&=&T_c\sigma_{x}(\cosh{A}-\eta)+T_ci\sigma_y(\sinh{A}-
{\eta}A),
\end{eqnarray}
where
\begin{eqnarray}
A=\sum_{\bf{q}}\frac{\xi_{\bf{q}}}{\omega_{\bf{q}}}(M_{\bf{q}}b^{\dag}_{\bf{q}}-M^{\ast}_{\bf{q}}b_{\bf{q}}),
\end{eqnarray}
and $\eta$ is a parameter which will be adjusted to minimize
perturbation terms ($H^{'}_{1}$ and $H^{'}_{2}$). Obviously,
$H^{'}_{0}$ can be solved exactly. We denote the ground state of
$H^{'}_{0}$ as
\begin{eqnarray}
&&|g\rangle=|s_{2}\rangle|\{{0_{\bf{q}}}\}\rangle,
\end{eqnarray}
and the lowest excited states as
 \begin{eqnarray}
&&|e_s\rangle=|s_{1}\rangle|\{{0_{\bf{q}}}\}\rangle,\\
&&|e_{\bf{q}}\rangle=|s_{2}\rangle|1_{\bf{q}}\rangle,
\end{eqnarray}
where $|s_{1}\rangle$ and $|s_{2}\rangle$ are eigenstates of
$\sigma_x$ ($ \sigma_{x}|s_{1}\rangle=|s_{1}\rangle$,
$\sigma_{x}|s_{2}\rangle=-|s_{2}\rangle$),
$|\{{0_{\bf{q}}}\}\rangle$ stands for the vacuum state for phonon,
and $|1_{\bf{q}}\rangle$ means that there is only 1 phonon for
mode $\bf{q}$ and no phonon for other modes. Let
$H^{'}_{1}|g\rangle=0$ and $ {\langle}g|H^{'}_{2}|g\rangle=0$, we
will get $\xi_{\bf{q}}$ and $\eta$ respectively as follows:
\begin{eqnarray}
&&\xi_{\bf{q}}=\frac{{\omega_{\bf{q}}}}{{\omega_{\bf{q}}}+2\eta{T_c}},\\
&&\eta=\exp[-\sum_{\bf{q}}\frac{|M_{\bf{q}}|^2}{2\omega_{\bf{q}}^2}\xi_{\bf{q}}^2].
\end{eqnarray}
Now we can easily check that ${\langle} e_s|H^{'}_1|e_{s}\rangle=0
$, ${\langle} e_{\bf{q}}|H^{'}_1|e_{\bf{q}}\rangle=0 $, ${\langle}
e_s|H^{'}_2|g\rangle=0 $, ${\langle} e_{\bf{q}}|H^{'}_2|g\rangle=0
$, and ${\langle} e_{\bf{q}}|H^{'}_1|e_s\rangle=V_{\bf{q}}$, where
$V_{\bf{q}}=2{\eta}T_c{M_{\bf{q}}\xi_{\bf{q}}/\omega_{\bf{q}}}$.
With these relations above, we can now diagonalize the lowest
excited states of $H^{'}$ as
\begin{eqnarray}
H^{'}&=&-{\eta}T_c{|}g{\rangle}{\langle}g{|}+\sum_EE{|}E{\rangle}{\langle}E|\nonumber\\
&~~& +{\rm terms~with~high~excited~states}.
\end{eqnarray}
The experiment in Ref. [13] is performed at lattice temperature
below 20 mK.\cite{13,14} At such a low temperature, the
multiphonon process is weak enough to be negligible. So we can get
the transformation as\cite{26,27,wuzj}
\begin{eqnarray}
&&|e_s\rangle =\sum_Ex(E){|}E{\rangle},\\
&&|e_{\bf{q}}\rangle=\sum_{E}y_{\bf{q}}(E){|}E{\rangle},\\
&&{|}E{\rangle}=x(E)|e_s\rangle+\sum_{\bf{q}}y_{\bf{q}}(E)|e_{\bf{q}}\rangle,
\end{eqnarray}
where
\begin{eqnarray}
&&|x(E)|^2=[1+\sum_{\bf{q}}\frac{|V_{\bf{q}}|^2}{(E+{\eta}T_c-\omega_{\bf{q}})^2}]^{-1},\\
&&|y_{\bf{q}}(E)|^2=\frac{|V_{\bf{q}}|^2}{(E+{\eta}T_c-\omega_{\bf{q}})^2}|x(E)|^2,\\
\end{eqnarray}
and the $E$'s are solutions to the equation
\begin{eqnarray}
&&E-{\eta}T_c-\sum_{\bf{q}}\frac{|V_{\bf{q}}|^2}{E+{\eta}T_c-\omega_{\bf{q}}}=0.
\end{eqnarray}

The population inversion can be defined as
$P(t)={\langle}\psi(t)|\sigma_z|\psi(t)\rangle$, where
$|\psi(t)\rangle$ is the total wavefunction (qubit and reservoir)
in Schr\"{o}dinger picture, and
\begin{eqnarray}
&&|\psi(t)\rangle=e^{-S}e^{-iH^{'}t}e^S|\psi(0)\rangle.
\end{eqnarray}
Since the qubit is initialized at the state $|L\rangle$, it is
reasonable to choose
$|\psi(0)\rangle=e^{-S}|L\rangle|\{{0_{\bf{q}}}\}\rangle$. Then we
can obtain
\begin{widetext}
\begin{eqnarray}
P(t)&=&{\langle}\{{0_{\bf{q}}}\}|{\langle}L|e^{iH^{'}t}e^S\sigma_{z}
e^{-S}e^{-iH^{'}t}|L\rangle|\{{0_{\bf{q}}}\}\rangle \nonumber\\
&=&-\frac{1}{2}\sum_E|x(E)|^2\exp{[-i(E+{\eta}T_c)t]}-\frac{1}{2}\sum_E|x(E)|^2\exp{[i(E+{\eta}T_c)t]}\nonumber\\
&=& -\frac{1}{4{\pi}i}\oint_Cd\omega e^{-i\omega
t}(\omega-2{\eta}T_c-\sum_{\bf{q}}\frac{|V_{\bf{q}}|^2}{\omega+i0^+-\omega_{\bf{q}}})^{-1}
\nonumber\\
&~~&-\frac{1}{4{\pi}i}\oint_{C^{'}}d\omega e^{i\omega
t}(\omega-2{\eta}T_c-\sum_{\bf{q}}\frac{|V_{\bf{q}}|^2}{\omega-i0^+-\omega_{\bf{q}}})^{-1},
\end{eqnarray}
\end{widetext}
where $\omega=E+{\eta}T_c$. Denoting the real and imaginary part
of $\sum_{\bf{q}}|V_{\bf{q}}|^2/(\omega\pm i0^+-\omega_{\bf{q}})$
as $R(\omega)$ and $\mp \gamma(\omega)$ respectively, we can get
\begin{eqnarray}
R(\omega)&=& \sum_{\bf{q}}
{\mathcal{P}} \frac{|V_{\bf{q}}|^2}{\omega-\omega_{\bf{q}}} \nonumber\\
&=& 4({\eta}T_c)^2 {\mathcal{P}} \int_0^\infty d
\omega^{'}\frac{J(\omega^{'})}{(\omega-\omega^{'})(\omega^{'}+2{\eta}T_c)^2},\\
\gamma(\omega)&=& \pi \sum_{\bf{q}} |V_{\bf{q}}|^2 \delta
(\omega-\omega_{\bf{q}}) \nonumber \\
&=&4\pi ({\eta}T_c)^2 \frac{J(\omega)}{(\omega +2 {\eta}T_c)^2},
\end{eqnarray}
where $\mathcal{P}$ stands for Cauchy principal value, and the
spectral density $J(\omega)$ is defined in Eq. (5). The parameter
$\eta$ determined by Eq. (15) and Eq. (16) can also be expressed
as
\begin{eqnarray}
&&\eta=\exp{[-\int_0^\infty
d\omega\frac{J(\omega)}{2(\omega+2{\eta}T_c)^2}]}.
\end{eqnarray}
The contour integral in Eq. (26) can proceed by calculating the
residue of integrand and the result is
\begin{eqnarray}
&&P(t)=-\cos(\omega_rt)\exp(-\gamma t),
\end{eqnarray}
where we have applied the second order approximation\cite{26}
\begin{eqnarray}
\gamma\simeq\gamma(2{\eta}T_c)=\frac{1}{4}{\pi}J(2{\eta}T_c),
\end{eqnarray}
and ${\omega_{r}}$ is the solution to the equation
\begin{eqnarray}
&&\omega-2{\eta}T_c-R(\omega)=0.
\end{eqnarray}
Finally, the tunneling electron population (probablity) in the
right dot at time t is given by
\begin{eqnarray}
n(t)=\frac{1}{2}[1+P(t)]=\frac{1}{2}[1-\cos(\omega_rt)\exp(-\gamma
t)].
\end{eqnarray}
 Thus a rather simple expression
for the dynamical tunneling is obtained analytically. The damped
oscillation indicated by this expression agrees with the
experiment in Ref. [13].

\section{Comparison with PER approach}

\begin{figure}[t]
  \includegraphics[width=1\columnwidth]{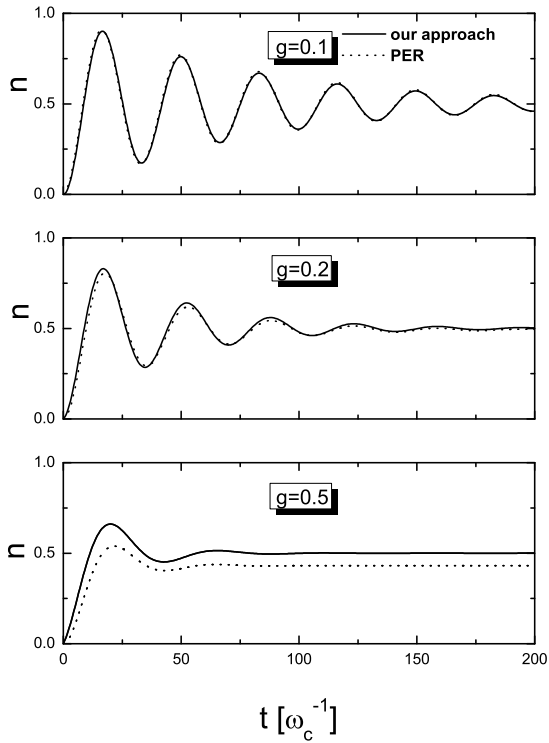}
\caption[]{\label{} The tunneling electron population in the right
dot as a function of time for Ohmic dissipation at low temperature
limit with the coupling constant $g=0.1$, 0.2, and 0.5. The
resonant tunneling rate is fixed as $T_c=0.1{\omega}_c$. The solid
curve and dot curve stand for the results of our approach and PER
approach respectively.\\
}
\end{figure}

\begin{figure}[t]
  \includegraphics[width=1\columnwidth]{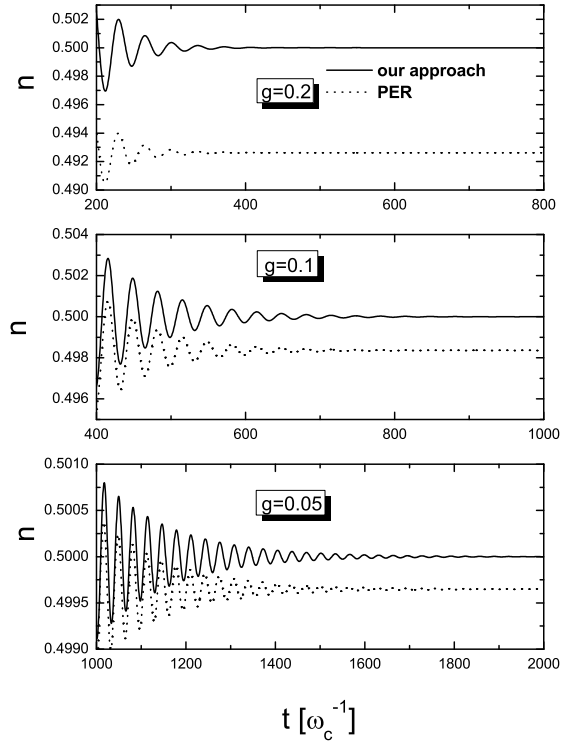}
\caption[]{\label{sidepeak.eps} Long time behavior of the
tunneling electron population in the right dot as a function of
time for Ohmic dissipation at low temperature limit with the
coupling constant $g$=0.2, 0.1, and 0.05. Others are the same as Fig. 1.\\
}
\end{figure}
As mentioned in the introduction above, Brandes has also derived
an expression for the dynamical tunneling in the double-dot
system. In his derivation the Born-Markov approximation is made
and the electron-phonon coupling $H_{ep}$ is treated as a
peterbation (PER).\cite{22} In this section, we give a comparison
between the PER and our approach.

The spectral density $J(\omega)$ defined in Eq. (5) is the only
quantity describing the interaction between the system and its
environment that enters into the dynamical tunneling. To get the
final result, we need the knowledge of this spectral function
first. Ordinarily, it can be written in a power law form with a
cutoff $\omega_c$,
\begin{eqnarray}
J(\omega)=g\omega^se^{-\omega/\omega_c},
\end{eqnarray}
where $0<s<1$, $s=1$, and $s>1$ corresponds to the sub-Ohmic,
Ohmic, and super-Ohmic spectral, respectively, g is a coupling
constant. Since the result for $n(t)$ is only presented for the
Ohmic case in Ref. [22], we also use the Ohmic spectral in this
section for convenience. We should point out that, our approach is
not limited by the form of $J(\omega)$. However, the Born-Markov
approximation made in PER approach is only meaningful and defined
for $s{\geq}1$.\cite{22}

Fig. 1 presents the time evolution of the tunneling electron
population (probability) in the right dot with three different
coupling constants. The result of both our approach and PER is
given at the temperature $T$=20 mK, at which the experiment is
performed.\cite{13} . The result at higher temperature can also be
obtained by PER and we find that the difference is
indistinguishable when the temperature is below 0.1 K, which
testifies the validity of the single phonon process assumption in
the transformation in Eq. (18), (19), and (20). The comparison
shows that, in weak coupling regime ($g\leq0.2$), the results of
two approaches agree with each other well; while in strong
coupling regime ($g\sim0.5$), the difference between two
approaches becomes clear. Although both two approaches are based
on perturbation theory, the perturbations in two approaches are
different. The PER approach treats the coupling $H_{ep}$ to the
phonon reservoir as a perturbation, which consequently requires
the coupling constant $g$ to be restricted in the weak coupling
regime. In our approach, the perturbation theory is exerted to the
Hamiltonian after a canonical transformation. The perturbation is
taken as $H^{'}_{1}$ and $H^{'}_{2}$, which can be minimized by
the variational parameter $\eta$. So our approach can be easily
extended to the strong coupling regime (it works well for the
whole range of $0<g<2$).\cite{26}

Then we focus on the long time behavior of the time evolution.
Since the double dot is symmetric (i.e., unbiased case,
$\varepsilon=0$), the population in two dots at long time limit
should also be symmetric. In other words, the asymptotic value of
$n(t)$ is expected to be 0.5. But in the strong coupling case
($g=0.5$) of Fig. 1, it has shown that there is a gap between the
result of PER at long time limit and that of our approach which
equals 0.5. Actually, even in weak coupling regime, this kind of
deviation still exists but rather small, as can be seen from Fig.
2 in an enlarged view for coupling constant $g=0.2$, 0.1, and
0.05. Since the asymptotic value from non-Markovian calculation
presented by DiVincenzo and Loss also equals 0.5,\cite{non-markov}
the deviation in PER approach must arise from Markov
approximation. And it is avoided in our approach.

There are two alternative ways of perturbation theory: one is in
the coupling $H_{ep}$ (PER), another is in the inter-dot coupling
$T_c$ (POL).\cite{22} To get a good approximation, one must
restrict the PER approach in the weak coupling regime, as analyzed
above. But it does not mean there is no requirement for $T_c$. In
Fig. 3, we present the dynamical population in the right dot
$n(t)$ in weak coupling ($g=0.1$), but with three different
int-dot tunneling rates $T_c=0.05,$ 0.01, and 0.001 ${\omega}_c$.
We find that, in strong inter-dot tunneling case
($T_c>0.05~{\omega}_c$), two approaches give consistent results;
while in weak tunneling case ($T_c<0.01~{\omega}_c$), the results
deviate from each other. It is well known that the energy
difference for the two eigenstates (i.e., the bonding and
anti-bonding states) of the double dot system is the Rabi
splitting $\Delta=\sqrt{\varepsilon^2+4T^2_c}$. Here in the
unbiased case ($\varepsilon=0$), the energy difference $\Delta$ is
determined only by the inter-dot tunneling $T_c$. When $T_c$ is
too small, the energy difference $\Delta$ between two states may
be smaller than the perturbation $H_{ep}$ of the PER approach. In
such situation, the energy states are nearly-degenerate and the
perturbation theory for a nondegenerate state of the PER approach
is not suitable. So the PER breaks down in weak inter-dot
tunneling regime. In our approach, however, the perturbation terms
($H^{'}_{1}$ and $H^{'}_{2}$) decrease with $T_c$, thus the
perturbation theory always works well for the whole range of
$T_c$.

\begin{figure}[t]
  \includegraphics[width=1\columnwidth]{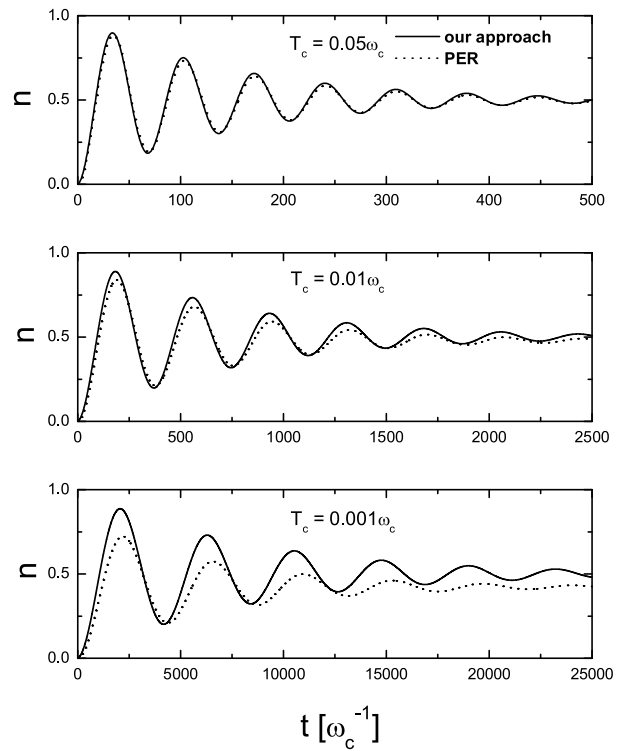}
\caption[]{\label{dlregime.eps} The tunneling electron population
in the right dot as a function of time
               for Ohmic dissipation at low temperature limit with the tunneling rate $T_c=0.05,$ 0.01, and $0.001{\omega}_c$.
               The coupling constant is fixed as
               $g=0.1$.}
\end{figure}

\section{Phonon induced decoherence}
One of the central points in quantum physics is the loss of
coherence of the quantum system. In this section, the spectral
density for the double-dot system coupling to bulk phonons is
derived from the microscopic view. Then the phonon induced
decoherence is analyzed at length. And the suppression of
decoherence is discussed.

\subsection{Spectral density}

In last section, we use the Ohmic spectral to calculate the
dynamical tunneling current. Obviously, it is not appropriate to
describe the properties of the double-dot system. Since all
relevant information of the environment and its coupling to the
qubit system is contained in the spectral function, it is
necessary and important to obtain a suitable spectral function to
describe a double quantum dot. Here, we only consider the coupling
to the bulk acoustic phonons, because we are interested in the low
temperature limit case. The only two types of interaction between
electrons and acoustic phonons in semiconductors are piezoelectric
coupling and deformation potential coupling.

We assume the electron wave functions ${\mid}L\rangle$ and
${\mid}R\rangle$ are sharply around the center of dot with
Gaussian shape $\sim$ exp$[-r^2/(2l^2)]$, where $l$ is the the dot
size. One can show that the piezoelectric coupling constant for
GaAs (zinc-blende structure) is\cite{20}
\begin{eqnarray}
M^{\rm{pz}}_{\bf{q},\lambda}&=&-(\frac{1}{2{\rho}qs_{\lambda}V})^{1/2}Me^{-a^2q^2/4}\nonumber\\
&~~&\times({\xi}^{\lambda}_1e_2e_3+{\xi}^{\lambda}_2e_1e_3+{\xi}^{\lambda}_3e_1e_2)\rm{sin}(\frac{\bf{q}\cdot\bf{d}}{2}),
\end{eqnarray}
where $\rho$ is the density of the crystal, $V$ is normalizing
volume, $s$ is the sound velocity in crystal (longitudinal sound
and transverse sound have different velocities), $e_i=q_i/q$,
$\bf{\xi}$ is the polarization vector whose components depend on
the polarization mode $\lambda$, $M$ is the piezoconstant, and $d$
is the center-to-center distance between two dots. Here, we
include the polarization again, since different polarization modes
give different coupling constants. With the simple dispersion
relation ${\omega}_{\bf{q},\lambda}=s_{\lambda}q$, one can now
calculate, according to the definition in Eq. (5), the spectral
function $J^{\rm{pz}}(\omega)$ due to piezoelectric coupling. But
the expression is rather complicated. To obtain a tractable form
of the piezoelectric coupling we use the angular average following
Bruus et al.\cite{28} and Brandes et al.\cite{15}. Then we get
\begin{eqnarray}
J^{\rm{pz}}(\omega)=g_{\rm{pz}}{\omega}(1-\frac{{{\omega}_{d}}}{\omega}{\rm{sin}}{\frac{\omega}{{\omega}_{d}}})
e^{-{\omega}^2/2{\omega_l}^2},
\end{eqnarray}
where $\omega_d=s/d$, $\omega_l=s/l$, and
$g_{\rm{pz}}=P/2{\pi}^2{\rho}s^3$ with
\begin{eqnarray}
P=M^2(\frac{12}{35}+\frac{1}{x}\frac{16}{35}),
\end{eqnarray}
where the transverse sound velocity is expressed as $x$ times the
longitudinal sound velocity $s$. For the deformation potential
coupling, the contribution from TA-acoustic phonons is small
enough to be neglected as compared with that from LA-acoustic
phonons. So the coupling constant can be written as\cite{20}
\begin{eqnarray}
M^{\rm{df}}_{\bf{q}}=iq{\Xi}(\frac{1}{2{\rho}qsV})^{1/2}
e^{-a^2q^2/4}\rm{sin}(\frac{\bf{q}\cdot\bf{d}}{2}),
\end{eqnarray}
where $\Xi$ is the deformation potential. Then we can easily get
the spectral function due to deformation coupling
\begin{eqnarray}
J^{\rm{df}}(\omega)=g_{\rm{df}}\omega^3(1-\frac{{\omega}_d}{\omega}{\rm{sin}}{\frac{\omega}{{\omega}_d}})
e^{-{\omega}^2/2{\omega_l}^2},
\end{eqnarray}
where $g_{\rm{df}}=\Xi^2/8\pi^2{\rho}s^5$.

With the parameters of GaAs in Ref. [30], we can estimate that
$g_{\rm{pz}}\approx0.035$ and
$g_{\rm{df}}\approx0.029~({\rm}ps)^{-2}$. Previous work states
that the contribution from deformation potential phonons is small
as compared with piezoelectric phonons in double-dot system of
GaAs material.\cite{22} Our result also proves it to be true in
the weak confinement regime (large dot size). But it is not valid
when the dot size is decreased to the strong confinement regime.
Fig. 4 shows the spectral functions $J^{\rm{pz}}(\omega)$ and
$J^{\rm{df}}(\omega)$ in strong confinement regime, with
$\omega_l=1~(\rm{ps})^{-1}$ (i.e., dot size $l=5~\rm{nm}$) and
$\omega_l=0.5~(\rm{ps})^{-1}$ (i.e., dot size $l=10~\rm{nm}$). As
we can see, $J^{\rm{df}}(\omega)$ is comparable to
$J^{\rm{pz}}(\omega)$ in that regime. But it shrinks much faster
than $J^{\rm{pz}}(\omega)$ as the dot size is increased and is
negligible when the dot size $l>50~\rm{nm}$. Fig. 4 also shows the
spectral functions $J^{\rm{pz}}(\omega)$ and $J^{\rm{df}}(\omega)$
at two different center-to-center distances, with
$\omega_d=0.05~(\rm{ps})^{-1}$ (i.e., $d=100~\rm{nm}$) and
$\omega_d=0.02~(\rm{ps})^{-1}$ (i.e., $d=250~\rm{nm}$). The
influence from the parameter $d$ to both $J^{\rm{pz}}(\omega)$ and
$J^{\rm{df}}(\omega)$ is small compared with that from the
parameter $l$. It adds the spectral function an oscillation term,
and the oscillation frequency (determined by $\omega_d$) is
increased with $d$. All these properties of spectral functions
$J^{\rm{pz}}(\omega)$ and $J^{\rm{df}}(\omega)$ determine the
decoherence induced by piezoelectric coupling and deformation
potential coupling to phonons, respectively.

\begin{figure}[t]
\includegraphics[width=1\columnwidth]{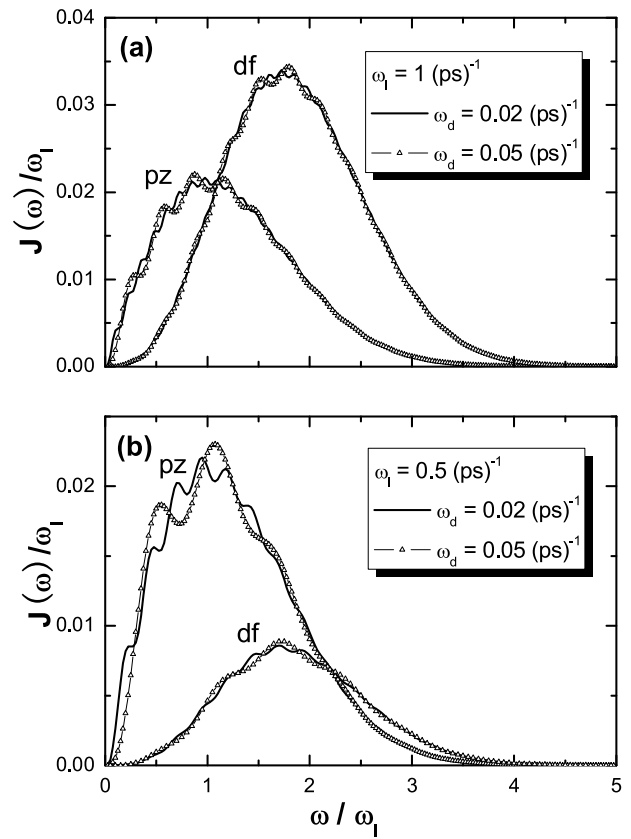}
\caption[]{\label{central.eps}Spectral functions of double quantum
dot due to piezoelectric coupling (labelled by pz) and deformation
potential coupling (labelled by df) with $\omega_d=0.02$ and
$\omega_d=0.05~(\rm{ps})^{-1}$. (a) $\omega_l=1~(\rm{ps})^{-1}$.
(b) $\omega_l=0.5~(\rm{ps})^{-1}$. }
\end{figure}

As one can see in Eq. (35) and Eq. (38), the deformation potential
coupling constant is real, while the piezoelectric coupling
constant is imaginary, which means they do not
interfere.\cite{28,29} Thus the total spectral density is
\begin{eqnarray}
J(\omega)=J^{\rm{pz}}(\omega)+J^{\rm{df}}(\omega).
\end{eqnarray}

\subsection{Decoherence induced by acoustic phonons}
The decoherence of quantum system due to interacting with
environment is a crucial point in quantum information. In a double
quantum dot, scattering by phonons can cause considerable loss of
coherence accompanied by dissipation when the tunneling electron
flips back and forth between two dots. One of the advantages of
our approach is that the decoherence rate in the this process is
obtained explicitly. Thus one can analyze it clearly.
\begin{figure}[t]
\includegraphics[width=1\columnwidth]{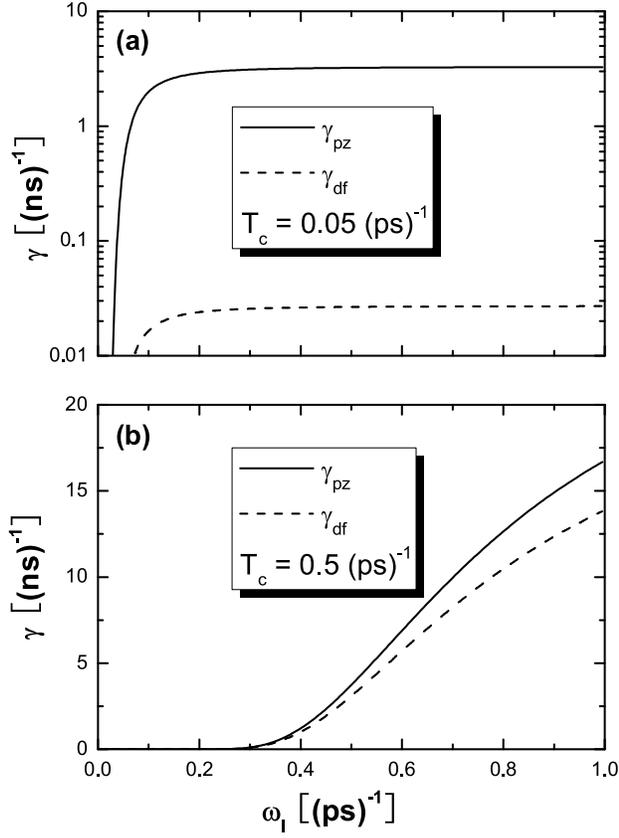}
\caption[]{\label{driving.eps}Decoherence rates
${\gamma}_{\rm{pz}}$ (solid line) and ${\gamma}_{\rm{df}}$ (dash
line) as functions of cutoff frequency $\omega_l$ when
$\omega_d=0.02~(\rm{ps})^{-1}$. The tunneling rates in Fig.5 (a)
and Fig.5 (b) are $0.05~(\rm{ps})^{-1}$ and $0.5~(\rm{ps})^{-1}$,
respectively.}
\end{figure}

Using the expressions of spectral density $J^{\rm{pz}}(\omega)$
(Eq. (36)) and $J^{\rm{df}}(\omega)$ (Eq. (39)) above, the
decoherence rates induced by piezoelectric and deformation
potential coupling are written as
\begin{eqnarray}
{\gamma}_{\rm{pz}}=\frac{1}{2}{\pi}g_{\rm{pz}}{\eta}T_c
(1-\frac{{{\omega}_{d}}}{2{\eta}T_c}{\rm{sin}}{\frac{2{\eta}T_c}{{\omega}_{d}}})
e^{-2{\eta}^2{T_c}^2/{\omega_l}^2},
\end{eqnarray}
and
\begin{eqnarray}
{\gamma}_{\rm{df}}=2{\pi}g_{\rm{df}}{\eta}^3{T_c}^3
(1-\frac{{{\omega}_{d}}}{2{\eta}T_c}{\rm{sin}}{\frac{2{\eta}T_c}{{\omega}_{d}}})
e^{-2{\eta}^2{T_c}^2/{\omega_l}^2},
\end{eqnarray}
respectively. Here, one should note the parameter $\eta$ in Eq.
(41) and that of the Eq. (42) are not the same, because they are
calculated from Eq. (29) with different spectral functions
($J^{\rm{pz}}(\omega)$ and $J^{\rm{df}}(\omega)$, respectively).
According to Eq. (40), the total decoherence rate induced by
acoustic phonons is
$\gamma={\gamma}_{\rm{pz}}+{\gamma}_{\rm{df}}$.

\begin{figure}[t]
\includegraphics[width=1\columnwidth]{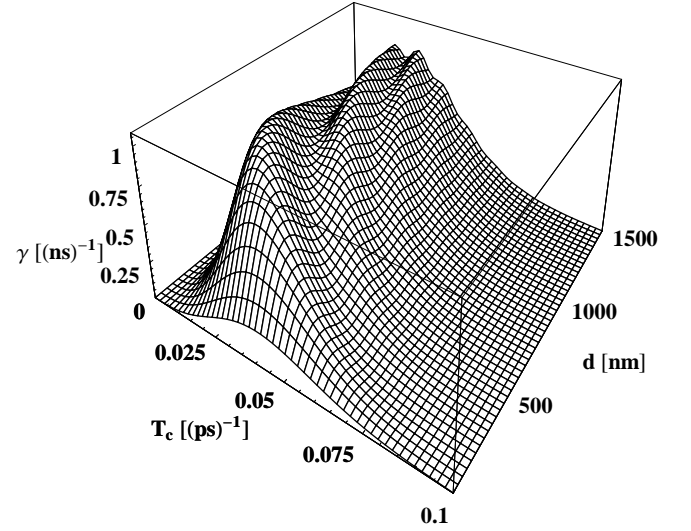}
\caption[]{\label{driving.eps}Decoherence rate $\gamma$ as a
function of tunneling rate $T_c$ and distance between two dots
$d$. The dot size is chosen to be $100~\rm{nm}$.}
\end{figure}

Fig. 5 presents the decoherence rates ${\gamma}_{\rm{pz}}$ and
${\gamma}_{\rm{df}}$ as functions of ${\omega}_l$ at two different
tunneling rates $T_c=0.05~(\rm{ps})^{-1}$ and
$T_c=0.5~(\rm{ps})^{-1}$. Another parameter $\omega_d$ is fixed as
$0.02~(\rm{ps})^{-1}$, which means the center-to-center distance
between two dots is about $250~\rm{nm}$.  As showed by Fig. 5(a),
at small tunneling rate, the contribution to decoherence rate
arose from deformation potential coupling is small compared with
that from piezoelectric coupling, even in strong confinement
regime (i.e., $\omega_l\sim1~(\rm{ps})^{-1}$). But when the
tunneling rate is large (Fig. 5(b)), the decoherence rates arising
from these two mechanisms are comparable, thus the contribution
from the deformation potential coupling can not be neglected. Fig.
5(a) and (b) also show that the decoherence rates (both
${\gamma}_{\rm{pz}}$ and ${\gamma}_{\rm{df}}$) are suppressed when
$\omega_c$ is decreased, which indicates that one should use large
dot size to get little decoherence. However, large dot size means
small characteristic energy spacing (on-site charging energy)
$\omega_{01}$ of single quantum dot. It is well known that, our
two level Hamitonian is valid to describe the double-dot system
only if $T_c$, $k_BT{\ll}\omega_{01}$, where $k_B$ is Boltzmann
constant.\cite{30} So, low temperature technique is needed to
maintain good quantum properties of dots when the dot size is
large, just as the experiment is performed.\cite{13}

In what follows, we choose a large dot size of $100~\rm{nm}$
(approximate size for the dot in Ref. [13]), i.e.,
$\omega_l=0.05~{(\rm{ps})^{-1}}$. Since the tunneling barriers in
experiment of Ref. [13] are made by depleting electrons with
negative gate voltage, their tunneling rates are
flexible.\cite{fujisawa} In Fig. 6, we present the decoherence
rate $\gamma$ ($\approx\gamma_{\rm{pz}}$ at that dot size) as a
function of tunneling rate and distance between two dots (from 100
to 1500 nm). Some oscillations, coming from the sine term in the
spectral density, are discerned from this 3-dimensional figure. We
find the characteristic decoherence time $T_2=1/\gamma$ speculated
from the figure is about $1~\rm{ns}$, which corresponds well with
the value fitted from the experimental curve.\cite{13} So the
coupling to phonons is one of the main decoherence mechanisms in
such a double-dot system. It also shows that the decoherence rate
increases with tunneling rate $T_c$ when $T_c<\omega_l$. But
larger tunneling rate will suppress the decoherence drastically,
even to zero when $T_c\approx0.1~(\rm{ps})^{-1}$ (i.e.,
$2\omega_l$). This value of $T_c$ is still in the range of
$\ll\omega_{01}=1.3~\rm{meV}\sim2~\rm(\rm{ps})^{-1}$,$^{13}$ in
which our two level model holds. Thus, such kind of decoupling
mechanism can be probably realized.

\section{Conclusion}
In conclusion, we have investigated the charge qubit dynamics in a
semiconductor double quantum dot coupled to phonons at low
temperature limit. Our approach is a perturbation theory after a
unitary transformation. The dynamical tunneling current is
obtained explicitly as a simple damped Rabi oscillation. The
comparison with PER approach shows the advantages of our approach
is that: it is not restricted by the form of spectral density; it
can be extended to strong coupling regime and works well for the
whole range of tunneling rate $T_c$; the long time behavior is
also consistent with the symmetric double-dot system.
Additionally, the simple decoherence rate expression allows us to
analyze the phonon induced decoherence clearly. We find that, in
strong confinement regime of dot and large tunneling rate $T_c$
($>0.1~\rm{(ps)}^{-1}$), the contribution to decoherence from
deformation potential coupling can not be neglected compared to
that from piezoelectric coupling in GaAs material. The decoherence
arose from both these two mechanism will be suppressed when the
dot size is increased. The decoupling with phonons will happen
when the condition $2\omega_l<T_c\ll\omega_{01}$ is met.

Finally, we hope our predictions can be testified by experiment in
the near future.

We thank T.Fujisawa for helpful discussions and careful reading of
the manuscript. This work was partly supported by National Natural
Science Foundation of China (No.10274051) and Shanghai Natural
Science Foundation (No.03ZR14060).


\end{document}